\documentclass[12pt]{article}
\usepackage{graphicx}

\usepackage{amsfonts,amsmath,amssymb,bm,eucal}
\usepackage{slashed}

\def\pbnr{BARI-TH/680-2013}
\def\speaker{Floriana Giannuzzi}
\def\title{New results in open charm spectroscopy through an effective Lagrangian approach}
\def\affiliation{Dipartimento di Fisica dell'Universit\`a di Bari and INFN, Sezione di Bari, Italy
}
\def\support{}

\newcommand\pubnumber{\pbnr}
\newcommand\pubdate{\today}
\def\Title#1{\begin{center} {\Large #1 } \end{center}}
\def\Author#1{\begin{center}{ \sc #1} \end{center}}

\newcommand{\OnBehalf}[1]{\sbox0{#1}\ifdim\wd0=0pt
        {}
	\else
	{\\on behalf of #1}
	\fi}
\newcommand{\SupportedBy}[1]{\sbox0{#1}\ifdim\wd0=0pt
        {}
	\else
	{\footnote{#1}}
	\fi}
\def\Address#1{\begin{center}{ \it #1} \end{center}}

\newcommand\pubblock{\includegraphics[width=5cm]{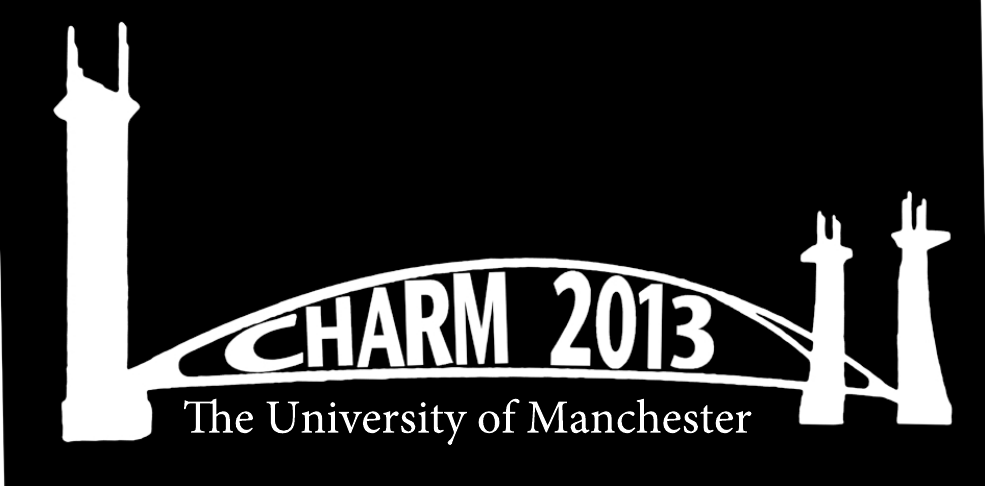}\hfill{\begin{tabular}{l} \pubnumber\\
         \pubdate  \end{tabular}}}
\newenvironment{Abstract}{\begin{quotation}  }{\end{quotation}}
\newenvironment{Presented}{\begin{quotation} \begin{center} 
             PRESENTED AT\end{center}\bigskip 
      \begin{center}\begin{large}}{\end{large}\end{center} \end{quotation}}
\def\Acknowledgements{\bigskip  \bigskip \begin{center} \begin{large}
             \bf ACKNOWLEDGEMENTS \end{large}\end{center}}
\def\venue{The 6$^{th}$ International Workshop on Charm Physics\\
(CHARM 2013)\\
Manchester, UK,  31 August -- 4 September, 2013}

\def\beq{\begin{equation}}
\def\eeq#1{\label{#1}\end{equation}}
\def\eeqn{\end{equation}}

\def\beqa{\begin{eqnarray}}
\def\eeqa#1{\label{#1}\end{eqnarray}}
\def\eeqan{\end{eqnarray}}

\let\bar=\overbar

\def\Dslash{\not{\hbox{\kern-4pt $D$}}}
\def\dslash{\not{\hbox{\kern-2pt $\del$}}}

\def\msb{{\bar{\ssstyle M \kern -1pt S}}}

\begin{document}
\begin{titlepage}
\pubblock

\vfill
\Title{\title}
\vfill
\Author{\speaker\SupportedBy{\support}}
\Address{\affiliation}
\vfill
\begin{Abstract}
An effective Lagrangian approach based on the heavy quark and chiral symmetry is introduced to analyse the spectroscopy of open charm mesons. Strong two-body decay widths and ratios of branching fractions are computed, and this piece of information is used to assign quantum numbers to recently observed charmed states which still need to be properly classified.
\end{Abstract}
\vfill
\begin{Presented}
\venue
\end{Presented}
\vfill
\end{titlepage}
\def\thefootnote{\fnsymbol{footnote}}
\setcounter{footnote}{0}
%

\section{Introduction}

In the last few years a big amount of data on heavy mesons, i.e. mesons comprising a light antiquark and a heavy quark (charm or beauty), has been generated from experiments.
The observed particles need a classification in terms of spin-parity, radial quantum number and angular momentum.
In the following, we will exploit the properties of QCD in the heavy quark limit to compute heavy-meson strong decay rates, focusing in particular on mesons with open charm; by comparing such quantities with available data from Belle and BaBar Collaborations, we will be able to propose or test their classification.
Recently, some analyses in this sector have also been performed by the LHCb Collaboration \cite{Aaij:2013sza}, and many states have been confirmed.

To determine the framework on which our results are based, it is worth reminding the properties of QCD in the heavy quark limit.
The heavy-quark QCD Lagrangian, in the limit $m_Q\to\infty$, has two symmetries: heavy-quark flavor symmetry, since it does not depend on the mass of the heavy quark, and heavy-quark spin symmetry, since it does not contain gamma matrices.
Standing the latter property, it turns out that the spin of the heavy quark ($s_Q$) and the total angular momentum of the light degrees of freedom ($\vec s_\ell=\vec \ell + \vec s_q$) are separetely conserved in strong interactions, and, as a consequence, heavy mesons can be arranged in doublets characterised by the value of $s_\ell$.
The total angular momentum of the two mesons belonging to the doublet $s_\ell$ is given by $J=s_\ell \pm 1/2$.
Such doublets can be described by effective fields, as summarized in Table~\ref{tab:doublets}.
Along with these states, we have also considered the first radial excitation ($n$=2), which will be denoted by putting a tilde over the symbol of the ground state.
\begin{table}[h]
\begin{center}
{\small
\begin{tabular}{lllll}
{\normalsize $\ell$} & {\normalsize $ s_\ell^P$ }& {\normalsize States} &\, {\normalsize $J^P$} \, & \,\,{\normalsize Doublets} \,\,\\
\hline
 0 & 1/2$^-$ & $(P,P^*)$ & $(0^-,1^-)$ & $ H_a=\frac{1+{\slashed{v}}}{2}[P_{a\mu}^*\gamma^\mu-P_a\gamma_5]$\\
\hline
1 & 1/2$^+$ & $(P_0^*,P_1^\prime)$ & $(0^+,1^+)$ & $S_a=\frac{1+{\slashed{v}}}{2} \left[P_{1a}^{\prime \mu}\gamma_\mu\gamma_5-P_{0a}^*\right]$\\
 & 3/2$^+$ & $(P_1,P_2^*)$ & $(1^+,2^+)$  & $T_a^\mu=\frac{1+{\slashed{v}}}{2} \left\{ P^{\mu\nu}_{2a}
\gamma_\nu - P_{1a\nu} \sqrt{\frac{3}{2}} \gamma_5 \left[ g^{\mu
\nu}-\frac{1}{3} \gamma^\nu (\gamma^\mu-v^\mu) \right]
\right\}$ \\
\hline
2 & 3/2$^-$ & $(P_1^*,P_2)$ & $(1^-,2^-)$ &  $X_a^\mu=\frac{1+{\slashed{v}}}{2} \left\{ P^{*\mu\nu}_{2a}
\gamma_5 \gamma_\nu -P^{\prime *}_{1a\nu} \sqrt{\frac{3}{2}}  \left[
g^{\mu \nu}-\frac{1}{3} \gamma^\nu (\gamma^\mu+v^\mu) \right]
\right\} $  \\
 & 5/2$^-$ & $(P_2^{\prime *},P_3)$ & $(2^-,3^-)$ & $X_a^{\prime\mu\nu}=\frac{1+{\slashed{v}}}{2} \left\{
P^{\mu\nu\sigma}_{3a} \gamma_\sigma -P^{*'\alpha\beta}_{2a}
\sqrt{\frac{5}{3}} \gamma_5 \left[ K^{\mu\nu}_{\alpha\beta} \right] \right\}$ \\
\end{tabular} 
}
\caption{Heavy meson doublets. In the $X^\prime$ expression, $K^{\mu\nu}_{\alpha\beta}=g^\mu_\alpha g^\nu_\beta - \frac{1}{5} \gamma_\alpha g^\nu_\beta (\gamma^\mu-v^\mu) -  \frac{1}{5} \gamma_\beta g^\mu_\alpha (\gamma^\nu-v^\nu)$.}
\label{tab:doublets}
\end{center}
\end{table}

We are interested in the decays $F\to HM$, of an excited meson ($F=\tilde H,S,T,X,X^\prime$) to the lowest lying doublet ($H$) plus a light pseudoscalar meson.
The Lagrangian terms describing these processes reproduce the properties of QCD in the heavy quark limit and respect chiral symmetry; they have been obtained at leading-order in the heavy-quark mass and in light meson momentum; the details about these Lagrangian terms and the expressions for the corresponding widths, that will be used in the following, can be found in \cite{Colangelo:2012xi}.
Since they depend on some effective parameters, the coupling constants, it is convenient to consider the ratios of widths in which such couplings cancel, in order to get  model independent quantities.
Notice that, since the momentum $\vec p$ of the light pseudoscalar meson in such decays is small, and since the widths scale as $|\vec p|^{2\ell +1}$ \cite{Colangelo:2012xi}, mesons decaying with a small (large) $\ell$ should be broad (narrow). This property will be as well used in the following discussion.

\section{Filling the doublets}
In this section, the classification of recently observed heavy mesons in the charm sector will be discussed. A summary of the results is presented, whereas the details about the computation can be found in \cite{Colangelo:2012xi}. 

\begin{description}
\item[$D_2^*(2460),D_{s2}^*(2573)$] ~\\
These states have $J^P=2^+$ ($T$ doublet) \cite{Beringer:1900zz}.
The comparison between our results and the experimental values \cite{Beringer:1900zz} of the following quantities confirms this assignement: ${\Gamma_1 \over \Gamma_2}={\Gamma(D^*_2(2460)^\pm \to D^0 \pi^+) \over \Gamma(D^*_2(2460)^\pm \to D^{*0} \pi^+) }=2.266 \pm 0.015$, $R_{12}={\Gamma_1 \over \Gamma_1 + \Gamma_2}=0.694 \pm 0.001 $  (to be compared with the experimental values ${\Gamma_1 \over \Gamma_2}=1.9 \pm 1.1 \pm 0.3$, $R_{12}=0.62  \pm 0.03 \pm 0.02$); ${\Gamma_1 \over \Gamma_2}={\Gamma(D^*_2(2460)^0 \to D^+ \pi^-) \over \Gamma(D^*_2(2460)^0 \to D^{*+} \pi^-) }=2.280 \pm 0.007$, $R_{12}={\Gamma_1 \over \Gamma_1 + \Gamma_2}=0.695 \pm 0.001 $ (to be compared with the experimental values ${\Gamma_1 \over \Gamma_2}=1.56 \pm 0.16 \pm 0.3$, $R_{12}=0.62 \pm 0.03 \pm 0.02$); ${{BR}(D^{*}_{s2}(2573) \to D^{*0} K^+) \over {BR}(D^{*}_{s2}(2573) \to D^0 K^+)}= 0.091 \pm 0.002$ (which should be lower than 0.33 according to experimental data).
The only discrepancy occurs in the neutral channel.

 \item[$D(2750),D(2760)$] ~\\
 They have been first observed by BaBar, and identified with the $\ell=2$ states  \cite{delAmoSanchez:2010vq}.
 Assuming they have $J^P=2^-,3^-$  ($X^\prime$ doublet), respectively, we find
 $ { {BR}(D^{*0}(2760) \to D^+ \pi^-) \over {BR}(D^{*0}(2750) \to D^{*+} \pi^-)}\Big|_{X^\prime \,{\rm doublet}}=0.660 \pm 0.001 $
 \cite{Colangelo:2012xi}, which agrees with the experimental measurement, giving $0.42 \pm 0.05 \pm 0.11$.

\item[$D_{sJ}(2860)$] ~\\
First observed by BaBar \cite{Aubert:2009ah}, it decays to both $DK$ and $D^*K$, so it can only be one of the following states: $1^-_{3/2},2^+_{1/2},3^-_{5/2}$. The first possibility is excluded by observing that it should decay in $p$-wave, and so should be a broad state, while the experimental width is $48\pm3\pm 6$ MeV. The quark model predicts a larger mass for the $2^+_{1/2} (n=2)$ state \cite{Di Pierro:2001uu}, so the $3^-_{5/2}$ assignment is supported \cite{Colangelo:2006rq}.
However, if we compare the experimental measurement ${BR(D_{sJ}(2860) \to D^*K) \over BR(D_{sJ}(2860) \to DK)}=1.10 \pm 0.15 \pm 0.19$ with the theoretical outcome, $0.39\pm0.01$, we find that it does not match the experimental number (this is also the case for the other possible assignements). Our claim is that, since its spin partner (whatever the classification is) is expected to have a similar mass (see its computation below), the experimental measurement may be contaminated by its decay. 
Indeed, if this contribution is added to the theoretical prediction, we find agreement and the identification with $3^-_{5/2}$ is saved \cite{Colangelo:2012xi}:
$ {\Gamma(D_{sJ}(2860) \to D^*K) +\Gamma(D_{s2}^{* \prime}(2851) \to D^*K)
\over \Gamma(D_{sJ}(2860) \to DK)} = 0.99 \pm 0.05 \,.$

\item[$D(2550),D^*(2600)$] ~\\
The Babar Collaboration has suggested that these states are the first radial excitation of the $\ell=0$ doublet \cite{delAmoSanchez:2010vq}. For $D^*(2600)$, the experimental ratio ${ {BR}(D^{*0}(2600) \to D^+ \pi^-) \over {BR}(D^{*0}(2600) \to D^{*+} \pi^-)}=0.32 \pm 0.02 \pm 0.09$ is not reproduced by the theoretical outcome  $0.822 \pm 0.003$.
Even assuming a different classification for $D^*(2600)$ among the allowed ones, this number cannot be reproduced by a state with such a mass \cite{Colangelo:2012xi}.
New observations by LHCb may shed light on this issue.

\item[$D_{s1}^*(2700)$] ~\\
This state with strangeness has been observed by Belle \cite{Brodzicka:2007aa} and BaBar \cite{Aubert:2006mh} both in $DK$ and $D^*K$ final state, so it has natural parity.
We support the hypothesis that it is the first radial excitation of $D_s(2112)$ ($1^-$) \cite{Colangelo:2007ds}, as confirmed by the comparison between the experimental measurement ${BR(D_{s1}^{*}(2700) \to D^*K) \over BR(D_{s1}^{*}(2700) \to
DK)} = 0.91 \pm 0.13 \pm 0.12$ \cite{Aubert:2009ah}, and the corresponding theoretical outcome in the present framework, being $0.91 \pm 0.03$.

\end{description}

Within this framework, the masses of not yet observed states can be predicted as well. To this aim, we will assume that the mass of the strange quark affects the mass of both mesons in a given doublet in the same way, or, in other words, the only effect of the mass of the strange quark is to shift the mass of the mesons in a given doublet by the same amount.
For instance, we can predict the mass of the spin partner of $D_{sJ}(2860)$, i.e. the state $D_{s2}^{\prime *}$, by requiring
$ M_{D_{s2}^{\prime *}}-M_{D_2^{\prime *}}=M_{D_{s3}}-M_{D_3}\,;$
we only assume that the non-strange states $D_2^{\prime *}=D(2750)$ and $D_3=D(2760)$ fill the same doublet in the $c\bar q$ spectrum as $D_{s2}^{\prime *}$ and $D_{sJ}(2860)$ in the $c\bar s$ spectrum, without specifying which doublet \footnote{A possible classification of $D(2750)$, $D(2760)$, $D_{sJ}(2860)$ in the $\tilde T$ doublet is discussed in \cite{Colangelo:2012xi}}.
Then, we get $M_{D_{s2}^{\prime*}}=2851 \pm 7$ MeV.
This value has been already used above in the discussion about $D_{sJ}(2860)$.

Analogously, we can compute the mass of $\tilde D_s$, the first radial excitation of the $0^-$ meson, by requiring
$ M_{D_s}-M_{D^0}=M_{\tilde D_s}-M_{\tilde D^0}\,,$
thus obtaining $M_{\tilde D_s}=2643 \pm 8$ MeV.
By writing a similar relation for the $1^-$ states,
we get $M_{\tilde D^{*0}}=2604 \pm 9$ MeV, a value which is compatible with the mass of $D^*(2600)$ and would confirm the identification of this state as the first radial excitation of the $1^-$ meson in the $c\bar q$ spectrum.

\section{Remarks and further applications}
\begin{figure}[htb]
\centering
\includegraphics[width=7.2cm]{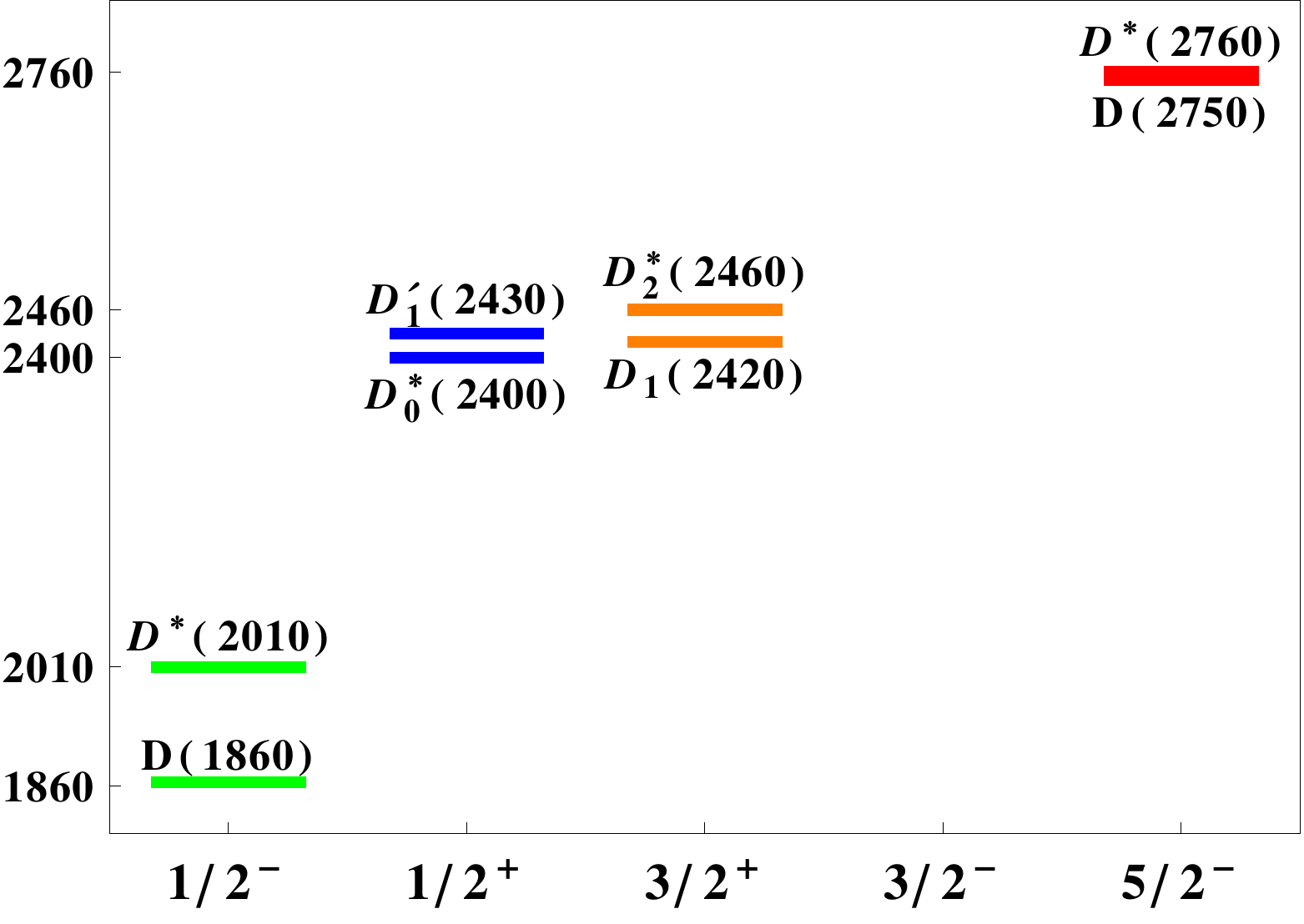}
\hspace*{.2cm}
\includegraphics[width=7.2cm]{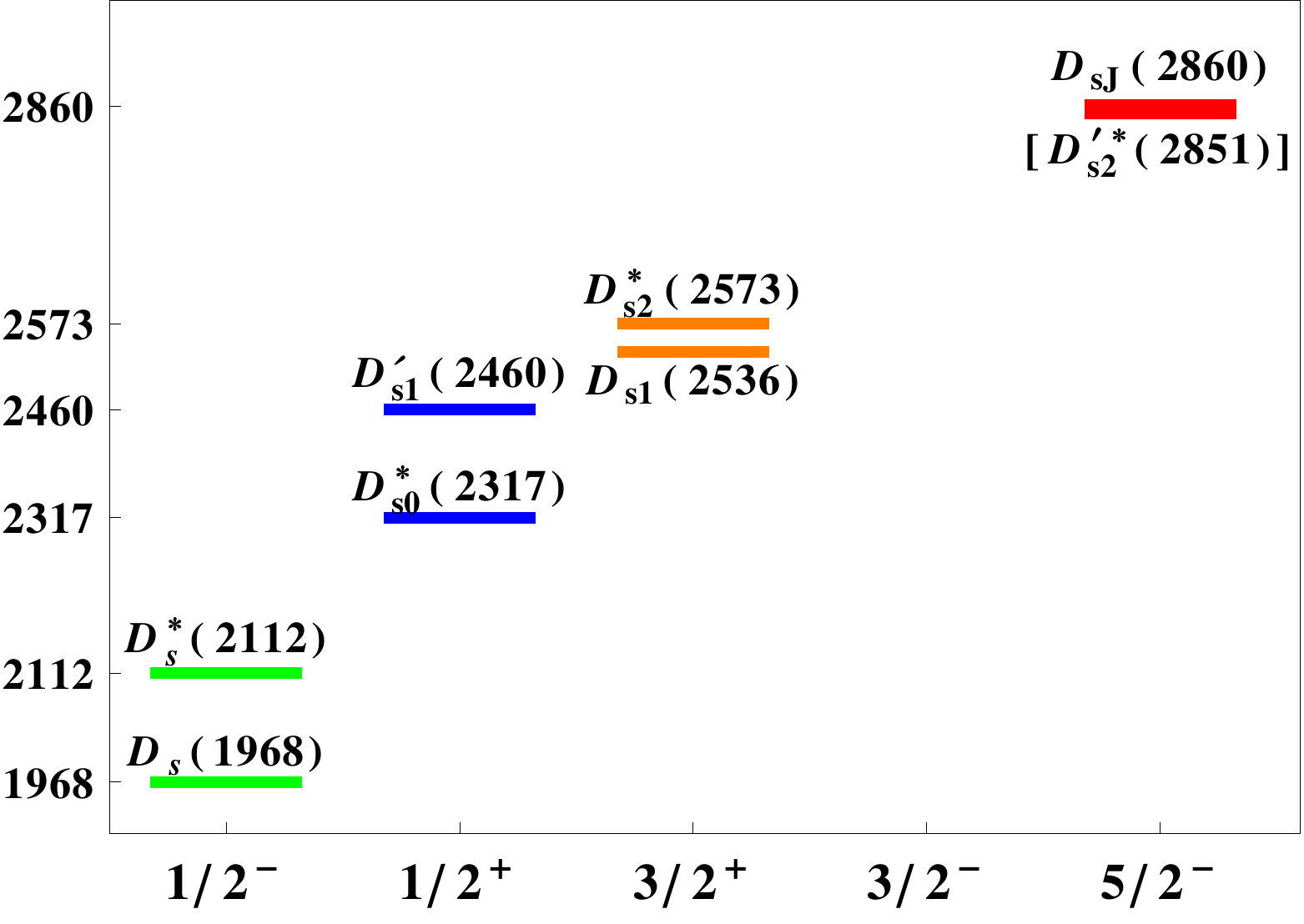}
\caption{$c\bar q$ (left panel) and $c\bar s$ (right panel) spectra, for $n$=1. The state $D_{s2}^{\prime *}(2851)$ has not been observed; its mass has been predicted within this framework.}
\label{fig:spectra}
\end{figure}

As a result, the $c\bar q$ and $c\bar s$ spectra would look as shown in Fig.~\ref{fig:spectra} (further experimental and theoretical details can be found in \cite{Swanson:2006st}).
The fundamental $1/2^-$ mesons, and the $1/2^+$ and $3/2^+$ ones are well established in both sectors.
Some issues remain for the $D_{s0}^*(2317)$ and $D_{s1}^\prime(2460)$: they have almost the same mass as their non-strange partners and are very narrow, while the $1/2^+$ states are expected to have a large width, their decay occurring in $s$-wave  \cite{Colangelo:2003vg}.
A discussion on the mass of $D_{s0}^*(2317)$ can be found in \cite{Mohler:2013rwa}, where a combined basis of quark-antiquark and $DK$ molecular operators has been used.
Furthermore, their small width can be understood by observing that they are below the $DK$ and $D^* K$ thresholds, respectively, so they can only have isospin violating decays.

Other applications of the framework discussed here regard the spectra of mesons with open beauty.
In order to estimate the mass of unknown beauty mesons, one can use charm data and take advantage of heavy-flavor symmetry, requiring
\begin{equation}\label{eq:deltalambda}
 \Delta_F^{(c)}=\Delta_F^{(b)} \,,\,  \lambda_F^{(c)}=\lambda_F^{(b)} \,,
\end{equation}
where $\Delta_F$ is defined as
$ \Delta_F=\bar M_F-\bar M_H\,,$
 $\bar M_{F}$ being the spin-averaged mass of a doublet $F$, and $\lambda_F$ accounts for the hyperfine splitting, i.e. the mass splitting between the two states in a doublet. 
 They appear in the first-order Lagrangian terms
\begin{equation}
 {\cal L}={1 \over 2 m_{Q}} \sum_F  \lambda_F \mbox{Tr} [{\bar
F}^{(\alpha)(\beta)}_{a} \sigma^{\mu \nu} F_{a(\alpha)(\beta)} \sigma_{\mu \nu}] \,,
\end{equation}
where $F$ represents a doublet in Table \ref{tab:doublets}, and are given by
\begin{equation}
\lambda_{H/S} = {1 \over 8} \left( M_{P^*}^2-M_P^2 \right) \quad
\lambda_{T/X}= {3 \over16} \left( M_{P^*}^2-M_{P}^2 \right) \quad
\lambda_{X^\prime}= {5 \over24} \left( M_{P^*}^2-M_{P}^2 \right)\,.
\end{equation}
Eq. \eqref{eq:deltalambda} means that the mass splittings $\Delta_F$ between doublets are the same for charmed and beauty mesons (first relation), and the mass splitting $\lambda_F$ between spin partners are the same  for charmed and beauty mesons (second one).
The l.h.s. of the two equations are fixed by experimental data; then, predictions for beauty mesons in the r.h.s. can be obtained.
Finally, it is also possible to fix the values of the coupling constants, by using the experimental data of the decay width.
A detailed discussion on these issues can be found in \cite{Colangelo:2012xi}.



\Acknowledgements
I am grateful to P. Colangelo, F. De Fazio and S. Nicotri for collaboration.

\end{document}